\documentstyle[12pt,epsf]{article}
\newcommand{\beq}{\begin{equation}}
\newcommand{\eeq}{\end{equation}}

\title{ Einstein's Boxes: 
\\Quantum Mechanical Solution  }

\author{  E.Yu.Bunkova, O.A.Khrustalev and O.D. Timofeevskaya 
\\ {\em  Moscow State University
(MSU) } \\ {\em Moscow, 119992, Russia.}
\\ {\it e-mail: olga@goa.bog.msu.ru}}

\date{ \ \ \  }
\begin{document}
\maketitle
\begin{abstract}
In this paper we present a solution of the Einstein's boxes
paradox by modern Quantum Mechanics in which a notion of density
matrix is equivalent to a notion of a quantum state of a system.
We use a secondary quantization formalism  in the 
attempt to make a description particularly clear. The aim of this paper 
 is to provide pedagogical help to the students  of quantum 
 mechanics. 
 \end{abstract}

\vspace{0.5cm}


\vspace{1cm}

\section{Introduction}

The "Einstein's Boxes" thought experiment is originally presented by Einstein
in 1927 at Solvay conference to demonstrate the incompleteness
in quantum mechanics description of reality. Later it was discussed and
modified by Einstein, de Broglie, Schrodinger, Heisenberg, and others,
into a simple scenario involving the splitting in half of the wave
function in the box. 

In his book \cite{Bro} de Broglie describes it in this way:

{\it
Suppose a particle is enclose in a box B with impermeable walls. The
associated wave $\Psi$ is confined to the box and cannot leave it. The
usual interpretation assert that the particle is "potentially" present
in the whole box $B$, with a probability $|\Psi |^2$ at each point.
Let us suppose that by some process or other, for example, by
inserting a partition into the box, the box $B$ is divided into two
separate parts $B_1$ and $B_2$ and that $B_1$ and $B_2$ are then 
transported to two very distant places, for example to Paris and Tokyo.
The particle, which has not yet appeared, thus remains potentially
present in the assembly of two boxes and its wave function $\Psi$ consists 
of two parts, one of which, ${\Psi}_1$, is located in $B_1$ and the other,
${\Psi}_2$, in $B_2$. The wave function is thus of the form $\Psi =
c_1{\Psi}_1 + c_2{\Psi}_2$, where $|c_1 |^2 +| c_2 |^2 =1.$

The probability laws of wave mechanics now tell us that  if an
experiment is carried out in box $B_1$ in Paris, which will enable
the presence of the particle to be revealed in this box, the 
probability  of this experiment giving a positive result is $|c_1 |^2$,
whilst the probability of it giving a negative result is $|c_2 |^2$.
According to the usual interpretation, this would have the following
significance: since the particle is present in the assembly of the two 
boxes prior to the observable, it would be immediately localized in the
box $B_1$ in the case of a positive result in Paris. This does not
seem to me to be acceptable. The only reasonable interpretation appears to 
me to be that prior to the observable localization in $B_1$, we know
that the particle was in one of the boxes $B_1$ and $B_2$, but we do
not know in which one, and the probabilities considered in  the usual
wave mechanics are the consequence of this partial ignorance. If we
show that the particle is in the box $B_1$, it implies simply that it was 
already there prior to localization...

We might note here how the usual interpretation leads to paradox in the 
case of
experiment with negative result...if
nothing is observed, this negative result will signify that the particle 
is not in  box $B_2$ and it is thus in box $B_1$ in Paris. But this
can reasonably signify only one thing: the particle was already in Paris 
in box $B_1$ prior to the drainage experiment made in Tokyo in box $B_2$.
Every other interpretation is absurd. How can we imagine that the simple 
fact of having observed {\bf nothing} in Tokyo has been able to promote 
the localization of the particle at a distance of many thousands of miles 
away?}

In  paper published recently, T. Norsen \cite{Nor} presents the history of 
the problem, several formulation of this thought experiment, analyses and 
assess it from point of Einstein-Bohr debates, EPR dilemma, and Bell's 
theorem. This paper has encouraged us to consider the 
problem of two boxes in the frame of modern quantum mechanics. 

The definition of 
quantum state of the quantum system was done by von Neumann \cite{Neu}
in 1927. The concept of density matrix's operator solves EPR-paradox
\cite{Scu} and other paradoxes in Quantum Mechanics. 

By means of Einstein's boxes problem we demonstrate the quantum mechanical
approach to description of quantum state of subsystem of complicate quantum 
system. The paper is organized as follows. In section $2$ we recall the
von Neumann formalism of Density Matrix operator and discuss the problem
of description of the state of subsystems of composite system with the 
help of reduced and conditional density matrices. In section 3 we formulate 
the problem in the boxes in secondary quantization formalism. In section 
4 we define the quantum states for each  box after separation. 
In section $5$ we consider the process of observation of the particle in the 
box $S_2$ and  can see that this process has no influence on the quantum 
state in the box $S_1$,   we also discuss the property of the
quantum  state of the particle while the boxes are conjugated again. In 
conclusion we review our consideration and ensure that there is not 
Einstein's boxes paradox.  
             
\section{  Density Matrix   }
\subsection{  Notion of Quantum State}
\noindent
The general definition of quantum state was given  by von Neumann \cite{Neu}. 

He proposed the following procedure for calculation of
average values of physical variables $\hat F$:
$$  < {\cal F} >
    \quad = \quad
        Tr({\hat F}{\hat {\rho}}).
$$

Here  operator $\hat \rho$  satisfies three conditions:
$$   1) \quad {\hat \rho}^{+} \quad = \quad
         {\hat \rho},
$$
$$   2) \quad Tr{\hat \rho} \quad = \quad 1,
$$
$$   3) \quad  \forall \psi \in {\cal H} \quad
   <\psi|{\hat \rho}\psi>  \quad \geq 0.
$$

By the formula for average values von Neumann found out the
correspondence  between linear operators $\hat \rho$ and states of
quantum systems.
This formula gives quantum
mechanical definition of the notion "a state of a system". The
operator $\hat \rho$ is called {\it Density Matrix}.

Suppose that  $\hat F$ is an operator with discrete non-degenerate
spectrum. If  an observable ${\cal F}$ 
has a definite value in the state  $\rho,$
i.e. a {\it dispersion} of ${\cal F}$  in the state $\hat \rho $ equals 
zero,  then  the density matrix of this state
 is a projective operator satisfying the condition
$$   {\hat \rho}^{2} = \hat \rho =\hat P_N ,
\quad \hat P_N=|{\Psi}_N
\rangle \langle {\Psi}_N |,\quad 
\langle {\Psi}_{N}|{\Psi}_{N} \rangle =1.
$$
The average value of an arbitrary variable in this state is equal
to
$$  \langle {\cal A} \rangle
     \quad = \quad
   \langle {\Psi}_{N}|{\hat A}{\Psi}_{N} \rangle .
$$
It is so-called {\it pure } state. If the state is not pure it is
known as {\it mixed.}

\subsection{Composite System and Reduced Density Matrix }
\noindent
Suppose     the system $S$ is an unification
of two subsystems $S_{1}$  and $S_{2}$:
$$   S \quad = \quad
S_{1} \cup S_{2}\,.
$$
Then  the Hilbert space $\cal H$ is
a direct product of two spaces  
$$  {\cal H} \quad = \quad {\cal H}_{1}\otimes{\cal H}_{2},
$$
here the  space  $\cal H$  corresponds to the system $S$ and the
spaces ${\cal H}_{1}$ and ${\cal H}_{2}$ correspond to the
subsystems $S_{1}$ and $S_{2}$.

Now suppose that a physical variable ${\cal F}^{(1)}$ is connected with
 subsystem $S_1$  only. The average value of this variable in the state 
$(\rho)_{1+2}$
is given by equation
$$ \langle {\cal F}^{(1)} \rangle_{\rho}
   \quad = \quad Tr(\hat F^{(1)}\hat{\rho}_{(1+2)} )\quad =\quad
Tr_1(\hat F^{(1)}\hat{\rho}_1 ),
$$
where the operator $\hat{\rho}_{1}$ is defined by the formula
$$   {\hat \rho}_{1} \quad = \quad
      Tr_{2}{\hat \rho}_{1+2}.
$$
The operator ${\hat \rho}_{1}$  {\cite{Ech}} satisfies all the properties of
Density Matrix in $S_1$. 
The operator
 is called  {\it Reduced Density Matrix} . Thus,
the state of the subsystem $S_1$ is defined by reduced density
matrix.
The reduced density matrix for the subsystem $S_2$ is defined
analogously:
$$   {\hat \rho}_{2} \quad = \quad
      Tr_{1}{\hat \rho}_{1+2}.
$$
Quantum states  $\rho_{1}$ and $\rho_{2}$ of subsystems are
defined uniquely by the state $\rho_{1+2}$ of the composite
system.

\subsection{Conditional Density Matrix}
\noindent The average value of a variable $\hat F^{(1)}\hat P^{(2)}$,
where $\hat P^{(2)}=|{\phi}^{(2)}\rangle _2 \langle {\phi}^{(2)}|$, 
$\langle {\phi}^{(2)}|{\phi}^{(2)}\rangle _2 =1, $ in the state $\hat\rho$ is
equal to
 $$  \langle F^{(c)} \rangle_{\rho}
   \quad = \quad Tr_1 (\hat F^{(1)}\hat {\rho}_1^{c}).
$$
It is easy to demonstrate that the operator
$$
\hat {\rho}_1^{c} = {1 \over p}  Tr_2(\hat \rho\hat P^{(2)})
$$
satisfies all the properties of density matrix in the space ${\cal H}_1$. 
Here $p$ is a probability to find a subsystem $S_2$ in the {\it pure}
state $|{\phi}^{(2)} \rangle _2$
$$
p \quad = \quad Tr_{(1+2)}(\hat \rho\hat P^{(2)}) \quad = \quad 
Tr_{2}(\hat \rho _2\hat P^{(2)}).
$$
So the operator
$\hat\rho _1^c $ defines some state of the subsystem $S_1$. 
This state called a {\it Conditional Density Matrix} \cite{Bel}.

It is density matrix for the subsystem $S_1$ under condition that
the subsystem $S_2$ is selected in the pure state $
\hat P_2  =|{\phi}^{(2)}\rangle _2 \langle {\phi}^{(2)}|$.

\section {Pure States in  Boxes}
\subsection {One Box}
\noindent
 Suppose  a particle  is  in a pure state with the
wave function $\Psi (x)$ in the box $S'$. 
It means that there exists a physical variable 
$\hat F$ with nondegenerate discrete spectrum that has a definite value in 
this state, i.e. a dispersion of this variable equals zero. We can write 
$\Psi (x)$ in the form:
$$
\Psi (x) = \alpha {\Psi}_1 (x) +\beta {\Psi}_2 (x) , \quad 
\int  |\Psi (x)|^2 dx =1,
$$
where 
$$|\alpha |^2 +|\beta |^2 =1,$$
$$
{\Psi}_1 (x)=0 ( x>0), \qquad
   _{1}\langle \Psi |\Psi {\rangle}_1 =1,
$$
$$
{\Psi}_2 (x)=0 ( x<0), \qquad 
 _{2}\langle \Psi |\Psi {\rangle}_2 =1.
$$
For example, in the box
$$
V(x)=\{ { 0\quad |x|< a \atop {+\infty \quad |x|> a } }
$$
energy is nondegenerate 
$$E_n ={{\hbar}^2\over 2m}({\pi\over 2a })^2 n^2\qquad n=1,2,...$$
The stationary states with energies \cite{Fl} 
$E_{n} ={{\hbar}^2\over 2m}({\pi\over 2a })^2 n^2$ 
have the wave functions
$$
{\Psi}_n (x) = \sqrt {1\over a} \sin{({\pi n\over 2a}(x-a) )}, \qquad 
\int _{-a}^{+a} |\Psi (x)|^2 dx =1.
$$
If $n$ is  even then $\Psi (0)=0$.
 Suppose  a particle is in the state with $n=2k$.
\subsection {Two Boxes}
\noindent
Now we consider a quantum system of two boxes.
The wave functions in  separated box have to satisfy a condition
$\phi (0)=0$. The Hilbert space of the system $S=$ $(S_1$ $ U$ $S_2) $
is  ${\cal H}={\cal H}_1\otimes {\cal H}_2$ where
$$
{\cal H}_1 = \{ {\phi}_1 (x): {\phi}_1 (x)=0,\quad if\quad x>0,\quad
{\phi}_1 (-a)={\phi}_1 (0)=0\} ,
$$
$$
{\cal H}_2 = \{ {\phi}_2 (x): {\phi}_2 (x)=0,\quad if\quad x<0,\quad
{\phi}_2 (a)={\phi}_2 (0)=0\} .
$$
Later we will use  secondary quantization formalism:
$$ [\hat \Psi (x), \hat \Psi (y)]=0=[\hat {\Psi}^+ (x), \hat {\Psi}^+ (y)],
$$
$$
 [\hat \Psi (x), \hat {\Psi}^+ (y)]=\delta (x-y).
$$
Vacuum state is
$$
|0\rangle =|0{\rangle}_1 \otimes |0{\rangle }_2, \quad 
\hat\Psi (x)|0\rangle =0.
$$
The operators of particles numbers are
$$ 
\hat N =\hat N_1 +\hat N_2,
$$                    
$$
\hat N_1 =\int _{x<0}\hat {\Psi}^+(x) \hat \Psi (x)dx,
\qquad\hat N_2 =\int _{x>0}\hat {\Psi}^+(x) \hat \Psi (x)dx .
$$
One particle quantum mechanics corresponds to the sector with $N=1$.

The state with wave function ${\Psi}^{(1)} (x)$ in this representation is
$$
   |{\Psi}_1\rangle =\int _{x<0} dx {\Psi}^{(1)} (x)\hat {\Psi}^+ (x)  
|0{\rangle}_1 * |0{\rangle }_2  =|{\Psi}^{(1)}{\rangle}_1*|0{\rangle }_2 
$$
and the state with wave function ${\Psi}^{(2)} (x)$  is
$$
 |{\Psi}_2\rangle =
|0{\rangle}_1*\int _{x>0} dx {\Psi}^{(2)} (x)\hat {\Psi}^+ (x)  |0{\rangle }_2  =
|0{\rangle }_1*|{\Psi}^{(2)}{\rangle}_2 .
$$

If $n$ is even $( n=2k )$ then the wave function ${\Psi }_n,$ satisfies 
the condition $\Psi (0)=0$ and belongs to Hilbert space ${\cal H}$.  

The 
energy levels of the composite system $S$ are degenerate and there are 
  two basic orthonormal wave functions with the same energy.  To define a 
pure state of the system $S$ with definite energy we have to define 
additional quantum number. For example, let a parity be this variable:  
$$\hat P \psi (x)=\psi (-x).$$ Initial wave function ${\Psi}_n (x)$ is odd 
$$ \hat P \Psi (x)=\Psi (-x)=-\Psi (x).  $$ Another wave function with the 
same energy is even $$ \hat P {\Psi}_+ (x)={\Psi}_+ (-x)={\Psi}_+ 
(x)=|\Psi (x)|.$$ The variable $\hat F=\hat P\hat H$ with nondegerate
spectrum defines these wave functions uniquely.

Another basis is defined by energy and variable of number of particles in one
of the box, for example, in the box $S_2$  $(\hat N_1 
=1-\hat N_2)$:  $$ \{N_2=0,\quad E=E_{n}\}:\quad |{\Psi}^{(1)}{\rangle}_1 
|0{\rangle}_2 , $$ $$ \{N_2=1,\quad E=E_{n}\}:\quad |0{\rangle}_1 
|{\Psi}^{(2)}{\rangle}_2 .  $$ 
The wave function of the state ${\Psi}_{n=2k}$
 equals 
$$
|{\Psi}_n \rangle \quad =\quad\alpha  |{\Psi}^{(1)}_k{\rangle}_1 
|0{\rangle}_2 +
\beta |0{\rangle}_1 |{\Psi}^{(2)}_k{\rangle}_2 , \quad \alpha =-\beta={1\over
\sqrt 2}.
$$
 The average of coordinate in this state is
$$
<x>=\int _{-a} ^{+a} |{\Psi}_n (x)|^2  x dx , \quad \omega (x)={1\over a}
{\sin}^2{{\pi \over a}kx}.$$
The average of momentum is equal to
$$
<p>=\int _{-a}^{+a} |\tilde{\Psi}_{2k}(p)|^2  p dx , \quad \tilde\omega (p)=
{2k^2\pi a\over \hbar}{1\over ({p^2 a^2\over {\hbar}^2}-k^2{\pi}^2)^2}
\{
{{\cos}^2{pa/\hbar},\quad if\quad k \quad is \quad odd,\atop 
{\sin}^2{pa/\hbar},\quad if \quad k \quad is \quad even.}
$$
 Dispersion of energy in this state  equals  zero.

\section {Quantum States of Subsystems}
\noindent
Suppose, that ${\Psi}_n (x)$ is an initial wave function.
 We put  a partition  at the point $x=0$ carefully.

Hamiltonian of the system is changed. Now there exist two boxes with 
partition which were described in subsection 3.2.

The wave function ${\Psi}_n(x)\in{\cal H}$ does not change. Now it is
the composite system of two boxes in pure state $|{\Psi}_n\rangle$. 
This circumstance defines all quantum properties of the composite system and 
its subsystems. 

This
state has definite energy $E_n$. The density matrix of this pure state is
$$
\hat {\rho}_n =|{\Psi}_n\rangle\langle {\Psi}_n|=
|\alpha |^2 |{\Psi}^{(1)}{\rangle}_1\langle{\Psi}^{(1)}|\otimes
|0{\rangle}_2\langle 0|+ 
$$
$$\alpha \beta ^* |{\Psi}^{(1)}{\rangle}_1\langle 0|\otimes
|0{\rangle}_2\langle {\Psi}^{(2)} |   +\alpha ^* \beta
|0{\rangle}_1 \langle{\Psi}^{(1)}|\otimes
 |{\Psi}^{(2)}{\rangle}_2\langle 0 |
+|\beta |^2 |0{\rangle}_1\langle 0|\otimes |{\Psi}^{(2)}{\rangle}_2
\langle {\Psi}^{(2)} |.
$$
It is entangled state: whiles the composite system has a wave function,
the quantum states of each of the boxes $S_1$ and $S_2$ are mixed.

The reduced density matrix of the box $S_1$ is
$$
\hat {\rho}_1 =Tr_2  \hat \rho =               
|\alpha |^2 |{\Psi}^{(1)}{\rangle}_1\langle{\Psi}^{(1)}|+ 
|\beta |^2 |0{\rangle}_1\langle 0|.
$$
The box $S_1$ is empty with probability  $|\beta |^2$.  There is a particle
with wave function ${\Psi }_k^{(1)}$ in the box $S_1$ with probability
 $|\alpha |^2$.

The reduced density matrix of the box $S_2$ is
$$
\hat {\rho}_2 =Tr_1  \hat \rho =               
|\alpha |^2 |0{\rangle}_2 \langle 0 |+ |\beta |^2 |{\Psi}^{(2)}{\rangle}_2
\langle {\Psi}^{(2)} |.
$$
The box $S_2$ is empty with probability  $|\alpha |^2$.  There is a particle
with wave function ${\Psi }_k^{(2)}$ in the box $S_2$ with probability
 $|\beta |^2$. 

If the particle  in $S_2$ is selected in {\it pure}
state $P_2=|\phi\rangle\langle \phi|$ (independently how and when it is done)
the state of $S_1$ is $$\hat {\rho}_{1/2}^{(c)}={1\over p}\langle \phi |\hat
\rho |\phi\rangle ,$$
 where $p$ is the probability to find a system $S_2$ in the
state $|\phi\rangle$.

For example, if the particle is in the box $S_2$ ($N_2=1$) or the energy 
is equal to $E_k$ (the processes of measurement are different 
but they select the same pure state ${\Psi}_k^{(2)}$), then  the state of 
$S_1$ is  $\it pure$ state $|0{\rangle}_1$, i.e. there is no particle in 
the box $S_1$ and this event is definite with probability $p_1=1$.

For example, if  the box $S_2$ is empty ($N_2=0$) or the energy 
is equal to $0$, then  the state of 
$S_1$ is  $\it pure$ state $|{\Psi}_k^{(1)}{\rangle}_1$ and 
this event is definite with probability $p_1=1$.

In the next section we illustrate  these results.

\section {Quantum States of Subsystems (continuation)}
\noindent
Now we send the box  $S_1$ to Tokyo and the box $S_2$ to Paris.

Suppose that we change our mind and send our boxes back to Moscow.
We put away a partition  carefully and found that  the system $S'$
is in the pure state with the wave function ${\Psi}_{n}$.

 \subsection {Measurement in the Box  }
\noindent

If not,
suppose,  Alice in Paris decided to look into the box $S_2$ and to see
if the particle in it or not.  We describe this process in manner of
von Neumann measurement. 

The Alice's detector (subsystem $ S3 $)
 could be in two different positions: 
\noindent if detector  did not have registered the particle,
$$
|no > =\Big( {1\atop 0} \Big),
$$
 and if it  did,
$$
|yes > =\Big( {0\atop 1} \Big).$$
At the moment of time $t=0$ 
the composite system $\Gamma =S \oplus (S3) $
is in the pure  state   
$$
|\Theta\rangle =|{\Psi}_n\rangle \otimes \Big({1\atop 0}\Big).
$$
Since vacuum $|0{\rangle}_2$ is  the state without particles in the box $S_2$
$$\hat N_2 |0{\rangle}_2 =0$$
and
$$ \hat N_2 |{\Psi}^{(2)}{\rangle}_2=
|{\Psi}^{(2)}{\rangle}_2 ,$$
the Hamiltonian of interaction of the detector and the subsystem $S_2$ 
could be represented, for example, in the form
$$
\hat H_I = \gamma\hat N_2 \hat {\sigma}_1,\quad \hat {\sigma}_1=\Big( 
{0\atop 1} {1\atop 0}\Big) .  
$$ 
Unitary evolution of the  state of the 
system $(S\oplus S3)$ leads to 
$$ 
|{\Theta}(t)\rangle =\exp{\big( -i{\hat H_I t\over \hbar}\big)}
|{\Theta}\rangle  .
$$ 
It is well known that 
$$ e^{-i\hat{\sigma}_1 t\gamma\over \hbar 
}=\cos{\gamma t\over \hbar } -i\sin{\gamma t\over \hbar }\hat {\sigma}_1.  
$$ 
Suppose the duration of exposition is $t={\hbar \over \gamma}{\pi\over 
2}$ then $$ e^{-i\hat{\sigma}_1 t\gamma\over \hbar}=-i\hat {\sigma}_1 .  
$$ 
The wave function of the system $\Gamma$ at the moment $t$  
 is 
$$
|{\Theta}(t)\rangle\quad = 
\quad \alpha  |{\Psi}^{(1)}{\rangle}_1 |0{\rangle}_2 \Big({1\atop 0}\Big)-
i\beta |0{\rangle}_1 |{\Psi}^{(2)}{\rangle}_2 \Big({0\atop 1}\Big).
$$
The state of the composite system  $\Gamma$ rests pure under 
unitary evolution. The state of the system $S$ has changed. It is mixed now
with density matrix:
$$
{\rho}_S\quad =\quad 
|\alpha|^2 |{\Psi}^{(1)}{\rangle}_1\langle {\Psi}^{(1)}|
|0{\rangle}_2\langle 0 |+
|\beta|^2  |0{\rangle}_1\langle 0 |
|{\Psi}^{(2)}{\rangle}_2\langle{\Psi}^{(2)}|.
$$ 
We can  see that the quantum state of the subsystem $S_1$ has not changed 
: it is $\hat {\rho}_1$ again.
Thus there is no paradox: the observation of the subsystem $S_2$ in Paris
does not change the state of the subsystem $S_1$ in Tokyo.
  All measurements in the system
$S_1$  give the same results before and after observation in the box $S_2$

\subsection {Back to One Box }
\noindent
What happens if we send two boxes to Moscow and carefully put away a 
partition? Suppose we decide to measure the energy of the system $S'$.

With probability $W_{N=2k} =1/2$ the energy is equal to $E_{N=2k}$.
With probability $W_{N=2m} =0$ the energy is equal to $E_{N=2m}, m\not=k$.
With probability $W_{2l+1}, \quad l=0,1,2,...$ the energy equals 
$E_{(2l+1)}$ where 
$$ W_{2l+1}= 2\Big(\int^1_0 \sin{\pi k y}\cos{\pi y(l+1/2)}dy \Big) ^2.  
$$
$W_{2l+1}=0$ if $k$ and $l$ have the same parity.

For example, if $k=1$ then
$$ W_{2l+1}= {2\over
{{\pi}^2 [({l+1\over 2})^2-1]^2}}\sin ^2 {\pi (l+1)\over 2}.
$$
Distributions of coordinate and momentum do not change.

\subsection{Selected States}
If Alice decides to select the states such that there is no particle in 
the box $S_2$ then the quantum state of 
the composite system $S$ under condition that the detector is found in the state
$| yes\rangle$ is the pure state $|{\Psi}^{(1)} 
\rangle _1\otimes |0{\rangle }_2 $. It is defined by two quantum numbers :
$N_1=1$ and the energy in the system of two separated boxes $E_n$.

The state of the box $S_1$ is the quantum  state with 
conditional density matrix 
$$
 \hat {\rho}^c_1\quad =\quad _2\langle 0|\hat {\rho }_{S}|0{\rangle}_2
{1\over |\alpha |^2 } .  $$ 
It is pure 
state  of the particle in the box $S_1$ with wave
function $\Psi _k^{(1)} (x)$.     This pure state does not change during 
observation (measurement) in the box $S_2$. We can look into the box 
$S_1$ before or after looking into $S_2$ but  the particle is  always 
in the box $S_1$ if the box $S_2$ is empty.
This state is selected under condition that 
the particle isn't seen in the box $S_2$. 

Let us now put away a partition.
 The state is not the stationary state of the system: particle in the 
box $S_1\oplus $ "empty
box"  together. The energy distribution is 
$ W_{N}.$ It is not a paradox, it is a quantum logic

\section*{Conclusions}
\noindent
The reduced density matrix and conditional density matrix notions
resolve Einstein's boxes paradox. Quantum state of the composite system
defines the quantum states of the subsystems uniquely.

The initial {\it pure}
  state becomes the {\it mixed} state during observation. But the observation
in the box $S_2$ does not change the state of the box $S_1$.


\begin{thebibliography}{99}
\bibitem {Bro}  
L. de Brogle,
 {\it "The Current Interpretation of Wave Mechanics: A Critical
Study"}, Elssevier Publishing Company,  1964.

\bibitem{Nor} T.Norsen, {\it Einstein Boxes},
 arXive: quant-ph/0404016 (2004).

\bibitem {Neu}  
J.von Neumann,  G\"ott. Nach. pp.
1--57, 245 -- 272, {\bf 1927}. See, also, 
J. von Neumann,
 {\it "Mathematische Grundlagen der Quantenmechanik"}, (Berlin) 1, 1932.

\bibitem {Scu}
C.D.Candell and M.O.Scully, {\it Phys.Rep.} {\bf 43}, 488-508 (1978).

\bibitem {Ech}
V.V.Belokurov, O.A.Khrustalev, V.A.Sadovnichy, O.D.Timofeevskaya,
{\it Partucles and Nuclei, Lett.} {\bf 1}, 116 (2003).

\bibitem {Bel}
V.V.Belokurov, O.A.Khrustalev, V.A.Sadovnichy, O.D.Timofeevskaya
 in {\it System and subsystems in quantum communication},
(Proceedings of XXII Solvay Conference on Physics, Delphi Latin, 2001),
World Scientific Publising, 555, 2003.

\bibitem {Fl}
S.Flugge, {\it Practical Quantum Mechanics I}, Springer, 1971.



\end{thebibliography}
\end{document}